# Structural colors with embedded anti-counterfeit features fabricated by laser-based methods

Sascha Teutoburg-Weiss[a], Marcos Soldera[a,b*], Felix Bouchard[a], Joshua Kreß[c], Yana Vaynzof[c] and Andrés Fabián Lasagni[a,d]

[a] Institut für Fertigungstechnik, Technische Universität Dresden, George-Bähr-Str. 3c, 01069 Dresden, Germany

[b] PROBIEN-CONICET, Dto. de Electrotecnia, Universidad Nacional del Comahue, Buenos Aires 1400, Neuquén 8300, Argentina

[c] Integrated Centre for Applied Physics and Photonic Materials and Centre for Advancing Electronics Dresden (CFAED), Technical University of Dresden, Nöthnitzer Straße 61, 01187 Dresden, Germany

[d] Fraunhofer-Institut für Werkstoff- und Strahltechnik IWS, Winterbergstr. 28, 01277 Dresden, Germany

sascha.teutoburg-weiss@tu-dresden.de, marcos.soldera@tu-dresden.de*, felix.bouchard1@tu-dresden.de, joshua.kress@tu-dresden.de, yana.vaynzof@tu-dresden.de, andres_fabian.lasagni@tu-dresden.de

* Corresponding author

## Abstract

Structural colors can be induced on metals not only to fabricate logos or decorative motives but also to embed anti-counterfeit features for product protection. In this study, stainless steel (EN 1.4301) plates are colorized by growing a thin oxide layer using direct laser writing (DLW) and hidden anti-counterfeit measures are included on their surfaces by direct laser interference patterning (DLIP) processing. The periodic microstructures resulting from the DLIP treatment have a spatial period of 1 µm and act as relief diffraction gratings, featuring a characteristic diffraction pattern. These microstructures are not visible to the human eye but are easily detectable upon shining a coherent beam on the surface. Furthermore, the reflectance over the visible spectrum of the colorized surfaces with and without the DLIP microtexture is measured, giving low differences in the color perception following the so-called "CIE L*a*b*" color space. Finally, a demonstrator is fabricated, in which colorized fields with and without the security features are shown.

## 1   Introduction

Deliberately colorizing metal surfaces can serve different purposes, such as decoration and product or brand protection for anti-piracy measures. Usually, surfaces are colorized by depositing a coating with a pigment that is selected to absorb or reflect specific wavelength bands. Despite its broad use, the pigments can degrade over time due to UV exposure, heat, and chemical or mechanical abrasion [1–3]. As the deposition methods are relatively simple and well-established, the colorized features can be easily imitated and counterfeited. In contrast, surfaces with structural colors interact with incoming light through different physical mechanisms, such as thin film interference, diffraction by periodic microstructures, scattering by micro/nanoparticles, or excitation of plasmon resonances or plasmon polaritons [4–7]. As a consequence, more complex manufacturing methods are required to manufacture surfaces that can produce structural colorations. For example, physical vapor deposition (PVD) methods, such as thermal evaporation or sputtering, can be used to grow thin films of different materials with controlled thickness for thin film interference [8]; focused ion beam milling can be employed for fabricating periodic arrays of nano-holes in thin films for extraordinary optical transmission through the interference of surface plasmon polaritons [9]; or lithography-based methods, like photolithography or nano-imprint lithography, can be exploited for patterning relief diffraction gratings yielding a holographic coloration [10,11]. Consequently, the designed surfaces and the corresponding structural coloration are significantly harder to reproduce. Furthermore, the resistance to deterioration of the structural colors is dependent on the physical properties of the bulk material and its surface.

Laser-based methods have already been used in the past to generate structural colors [12–15]. The most widespread setup for this purpose is based on the direct laser writing (DLW) method, whereby a laser beam is focused onto the sample and scanned over the surface by galvanometer scanners [16]. As the mechanisms of laser-matter interaction depend strongly on the laser parameters and the material properties, entirely different surface modifications can occur for different processes. For instance, irradiating a metal surface with nanosecond (or longer) pulses, the material is allowed to heat up and react with the oxygen of the atmosphere, forming a continuous oxide film [17,18]. Adjusting the laser parameters, the film can grow to a desired thickness, typically on the order of the wavelength of visible light (~ 1 µm), so that thin film interference can take place. By fine-tuning this thickness, it is possible to obtain a colorful palette [19–21].

In addition, nanoparticles of noble metals, but also of some dielectrics, can produce strong resonance interactions with visible light, yielding vibrant structural colors. Varying the size of the nanoparticles, e.g. in the range 5 – 50 nm, different wavelengths can be scattered, tailoring the observable color [22]. Nanoparticles can be formed during a laser process with DLW when a short or ultra-short pulse (USP) ablates the surface of a material, increasing locally the temperature and creating a plasma plume. In the process, droplets of the material are ejected and rapidly condense and solidify [23,24]. Varying the cumulated fluence (total energy applied per unit of area), the shape and size of the nanoparticles can be controlled [25].

Periodic surface structures with periods of the order of the wavelength of visible light and up to a few microns can act as diffraction gratings for visible radiation. Currently, two laser-based methods stand out for this purpose due to their high throughput, namely laser-induced surface structuring (LIPSS) and direct laser interference patterning (DLIP), which can pattern thousands of micro or sub-microstructures with a single pulse. LIPSS are self-assembled structures with the shape of ripples, grooves, or cones that form upon the interaction of a metal or dielectric with a USP with a fluence close to the fluence threshold of the material [26,27]. The shape and alignment of the textures can be controlled with the cumulated fluence and polarization direction. Although the feature size can be modified by changing the incidence angle, the resulting sizes are normally on the order of the laser wavelength [28]. Following this approach, many authors have reported structural colors with a strong iridescent (viewing angle-dependent) appearance [29–32] as well as ultra-low light reflecting or black metals [33]. In turn, DLIP is based on the interference phenomenon arising upon overlapping two or more laser beams on the sample. At the maxima positions of the interference pattern, the material can be locally melted or ablated, whereas at the minima positions the surface remains unchanged. In this way, a periodic modulation of the topography, composition, oxidation state or crystallographic structure can be achieved [34–36]. The spatial period, typically in the range 500 nm – 20 µm, can be easily controlled by adjusting the overlapping angle and laser wavelength [37]. Furthermore, the texture shape and aspect ratio can be controlled by the number of overlapping beams, their polarization, and cumulated fluence [38]. Many works showing structural colors induced by DLIP on different materials have been reported [39–43].

The above-mentioned laser-based techniques share several advantages over other methods. They can produce features not only in the microscale but also below the diffraction limit, they are very flexible, in the sense that customized motives can be fabricated in individual workpieces without radical changes in the setup, and they do not require a pre-fabricated mask, cleanroom environments, photoresists or other chemical agents. As a consequence, laser-based methods offer a unique combination of features for high-throughput, cost-effective, and flexible fabrication of surfaces with structural colors. Furthermore, the different physical mechanisms governing the structural coloration described above can be combined on the same surface using laser-based manufacturing platforms, giving rise to complex and hard-to-imitate decorative features. This unexplored option determines the objective of study.

In this work, stainless steel plates were colorized by growing a thin oxide layer using DLW and hidden anti-counterfeit measures were included on their surfaces by DLIP processing. These features are not visible to the human eye but are easily detectable upon shining a coherent beam on the surface.

## 2 Materials and Methods

### 2.1 Materials

Stainless steel (EN 1.4301) plates supplied by SG Designbleche GmbH with dimensions of 100 x 100 mm² were laser-processed to obtain structural colors. The chemical composition of the steel plates is given in Table 1 according to the supplier. Before the laser structuring steps, the samples were electro-polished with a resulting surface roughness of Sa = 0.06 µm.

*Table 1. Chemical composition of the used stainless steel (EN 1.4301) plates.*

|      | C      | Mn   | Si   | P     | S      | Cr    | Ni   |
|------|--------|------|------|-------|--------|-------|------|
| wt%  | 0.0466 | 1.37 | 0.46 | 0.028 | 0.0006 | 18.07 | 8.11 |

### 2.2 Laser systems

To induce colors on the steel plates by thin film interference, a direct laser writing (DLW) system (GF Machining Solutions, Switzerland) equipped with a ytterbium fiber and a galvanometer scanner (Figure 1a, LS: laser scanner, S: sample) was used to heat the samples' surface and generate an oxide layer. This system allows for variable discrete pulse durations from 4 ns up to 200 ns with a maximum average power of 30 W and a frequency range of 2 - 1000 kHz. The nominal laser wavelength is 1064 nm with a beam quality $M^2$ < 1.5. The spot diameter at the focal plane is 60 µm.

Diffractive features were structured on the steel samples by direct laser interference patterning (DLIP). This structuring process was done with an Nd:YAG laser (NeoLase, Germany) at the doubled harmonic wavelength of 532 nm with a maximal average power of 7.5 W. The laser source provides a maximum pulse energy of 60 µJ at 10 kHz and 70 ps pulses. The system setup can be seen in Figure 1b, showing the optical head (DH: DLIP head) that enables the interference of four overlapping sub-beams on the sample surface. A schematic representation of the configuration and beams paths is shown in Figure 1e (DH). The primary beam from the laser source (L) is split by a diffractive optical element (DOE) into four equally intense sub-beams. These beams become parallel after refracting by the facets of a pyramidal prism (P). The distance between the sub-beams can be adjusted by the translation distance between DOE and prism, which in turn defines the spatial period of the interference pattern [44]. A fixed focus optical lens overlaps the sub-beams at the sample (S). For all the four-beams DLIP experiments a single pulse with an energy of 50 µJ, implying a fluence of 2.5 J/cm$^2$, at a repetition rate of 10 kHz was applied and the spatial period was fixed at 1.0 µm.

Both DLW and DLIP laser treatments were performed following the structuring strategy displayed in Figure 1d. Namely, the pulses were overlapped in the primary structuring direction (vertically in Figure 1d), either by the scanner system in the DLW machine or the mechanical axes in the DLIP machine, to generate a structured line featuring a width equal to the spot diameter. Then, the sample was shifted horizontally a hatch distance (H) and another vertical line was structured again. This sequence was repeated to cover the required area of the sample.

## 2.3 Characterization methods

Topographical data of the samples were acquired by confocal microscopy with an S Neox 3d Surface Profiler (Sensofar, Spain), with 50X and 150X objectives providing lateral resolutions of 170 nm and 140 nm and vertical resolutions of 5 nm and 1 nm, respectively.

To detect and record the diffraction patterns reflected from the periodic micropatterns, a Diffraction Measurement System (DMS), shown in Figure 1b, was developed at the TU Dresden and described in detail elsewhere [45]. This optical setup is mounted side-by-side to the DLIP head. Figure 1e (DMS) gives a schematic representation of its optical components and beam path. The emitted light from a low-power laser diode (L) is polarized by a rotational mounted linear polarizer to achieve maximum reflection by a polarization beam-splitter for optimum illumination of the sample surface. Through a lambda-quarter-plate, the polarization shifts first to circular before passing through a lens and illuminating the sample. After reflection on the sample, the light polarization shifts again back to linear by the lambda-plate but with a 90° offset permitting it to propagate through the beam-splitter. If periodic diffractive features exist on the samples´ surface, the diffraction orders of the reflected light will be captured with the optical configuration and guided to a CCD camera sensor. The acquired images can be utilized to further characterize the surface topography. Simulated results have shown that line-like periodic surface structures with periods from 10 µm down to 1.33 µm could be detected with this configuration [45]. For all the images collected by the CCD camera, an exposure time of 7 ms was set. The spot diameter at the sample surface was ca. 1 mm, ensuring that several tens of thousands of DLIP microstructures are illuminated simultaneously in order to produce well-defined diffraction modes.

The quantification of the structural colors was done by capturing the spectral reflectance of the samples with the setup shown in Figure 1c, featuring a halogen lamp (OceanOptics GmbH, HL-2000, Germany), a spectrometer (OceanOptics GmbH, HR2000+, Germany), and an integrating sphere (Thorlabs, IS200, USA). The sample was placed at the opening S of the integrating sphere (Figures 1c, f) at a tilting angle of 8° so that also the specular beam is collected by the integrating sphere as well. In addition, colorimetric values of the CIE L*a*b* color space were obtained from the reflectance data with the software OceanView (OceanOptics GmbH, Germany). The reference for the reflectance was a white reflection standard made of spectralon. The color space CIE L*a*b* was chosen over CIE xyz or other standards since it is closer to the human perception of color and therefore more suitable for the task of chromaticity comparison for human seen colors [46,47]. The values of this color space represent the lightness L* from 0 to 100, the position in the green (-100) to red (+100) range a*, and the position in the blue (-100) to yellow (+100) range b*. The CIE L*a*b* values measured with the illuminant type A were recalculated to CIE XYZ100 illuminant A, CIE XYZ100 illuminant D65, and finally CIE L*a*b* illuminant D65, where illuminant D65 represents the spectra of natural daylight. Functions and calculations were programmed with Python libraries [48,49].

X-ray photoemission spectroscopy (XPS) measurements were performed with a Thermofisher Escalab 250Xi system. All samples were measured in ultrahigh vacuum ($10^{-10}$ mbar). The x-rays were created from a monochromated AlKα source (1486.6 eV). For the measurements a pass energy of 20 eV was used. In order to remove surface contamination, the samples have been sputtered with monoatomic argon ions (3000 eV kinetic energy for 20 s) prior to the XPS measurements.

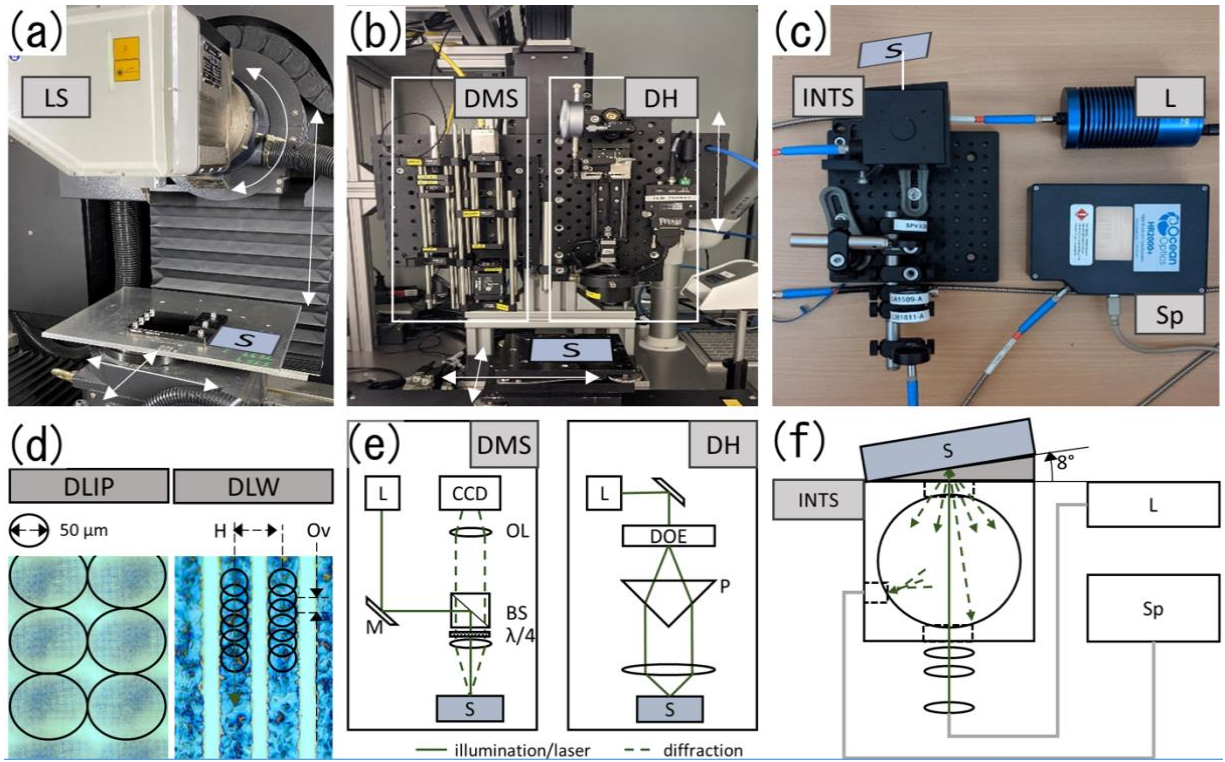

*Figure 1: (a) DLW system with scanner LS; (b) DLIP setup with DLIP optical head DH and diffraction measurement system DMS; (c) UV-vis photospectrometer experimental setup with integrating sphere INTS and spectrometer Sp; (d) structuring strategy for DLIP and DLW processes; (e) optical paths and components of DMS and DH and (f) integrating sphere-based spectrometer (S: sample, Ov: overlap, H: hatch, L: laser/light source, OL: optical lenses, M: mirror, BS: beam-splitter, DOE: diffractive optical element and P: pyramid).*

# 3 Results and Discussion

## 3.1 DLIP and DLW processing of stainless steel surfaces

The followed approach for combining colorized surfaces with DLIP structured features was based on growing an oxide layer by DLW on previously DLIP-structured surfaces, as indicated in Figure 2a. The underlying DLIP texture had an area of 80 × 80 mm², whereas the 12 different fields treated by DLW had an area of 15 × 40 mm², each covering featureless (i.e. without underlying DLIP texture) and DLIP-structured domains. Colorized areas on featureless steel surfaces were labeled as A to L, whereas colorized areas on the DLIP texture were labeled with an additional plus sign as A[+] to L[+] (see Figure 2b).

The four-beam DLIP treatment resulted in a hole-like microtexture with a spatial period of 1.0 µm and an average depth of 150 nm. According to previous investigations, this targeted topography was chosen as it yields high diffraction efficiencies [40,41].

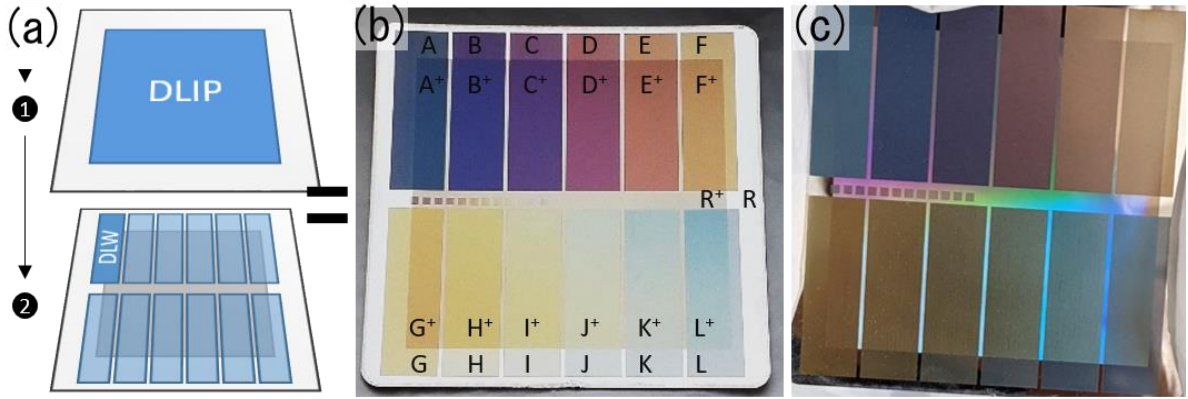

*Figure 2: Schematic of the two-step approach (a) with single full DLIP structured area (a.1) and subsequently DLW treated and colorized fields (a.2); photograph of the processed sample (b) with structural colors produced with different scanning speeds from 10 to 120 mm/s on featureless areas A to L and DLIP-structured areas A+ to L+. R and R+ are non-colorized references for featureless and DLIP-structured areas, respectively.*

To select the best DLW parameters, a parametric study with pulse energies from 50 µJ up to 1 mJ, scanning speeds from 5 to 200 mm/s, and pulse repetition rates from 30 to 300 kHz was conducted (not shown). The used pulse durations for this study were 4 ns and 200 ns, which are the minimum and maximum values that can be achieved with the employed laser system, respectively. Based on the perception of color saturation, color range, and general appearance of the resulting irradiated surfaces, all further DLW-based colorization experiments in this work were performed with the parameters shown in Table 2.

*Table 2: DLW parameters for colorizing stainless steel.*

| Parameter | Sample label | | | | | | | | | | | |
|---|---|---|---|---|---|---|---|---|---|---|---|---|
| | A | B | C | D | E | F | G | H | I | J | K | L |
| pulse duration [ns] | | | | | | -- 4 -- | | | | | | |
| pulse energy [µJ] | | | | | | -- 32 -- | | | | | | |
| fluence per pulse [J/cm²] | | | | | | -- 1.13 -- | | | | | | |
| peak fluence [J/cm²] | | | | | | -- 2.26 -- | | | | | | |
| repetition rate [kHz] | | | | | | -- 150 -- | | | | | | |
| scanning speed [mm/s] | 10 | 20 | 30 | 40 | 50 | 60 | 70 | 80 | 90 | 100 | 110 | 120 |
| hatch distance [µm] | 90 | 87 | 85 | 83 | 80 | 75 | 73 | 72 | 70 | 68 | 66 | 65 |
| cum. fluence [J/cm²] | 1019 | 509 | 340 | 255 | 204 | 170 | 146 | 127 | 113 | 102 | 93 | 85 |

To investigate the nature of the grown oxide layers, XPS measurements were performed on the selected samples A, D and G. The XPS spectra of (a) Fe $2p$ and (b) Cr $2p$ of the selected samples together with the corresponding fits are shown in Figure 3. Table S1 of the supplementary information lists all the identified species. For all samples, the spectra were fitted by the distinctive peaks corresponding to $Fe_2O_3$ and $Cr_2O_3$ and spinel compounds of the form $XYO_4$, where X and Y stand for ions with +2 and +3 valence. Following the reference [50], these spinel structures can be regarded as $(Mn^{2+}_{x1}Ni^{2+}_{x2}Fe^{2+}_{x3})(Fe^{3+}_{y1}Cr^{3+}_{y2})O_4$, where x1 + x2 + x3 = 1 and y1 + y2 = 2. Additionally, small amounts of NiO and MnO were found (see Table S1 of the supplementary information). All of the observed species were also detected by other works, in which steel substrates were deliberately oxidized by laser irradiation [35,50–53], further supporting our statement.

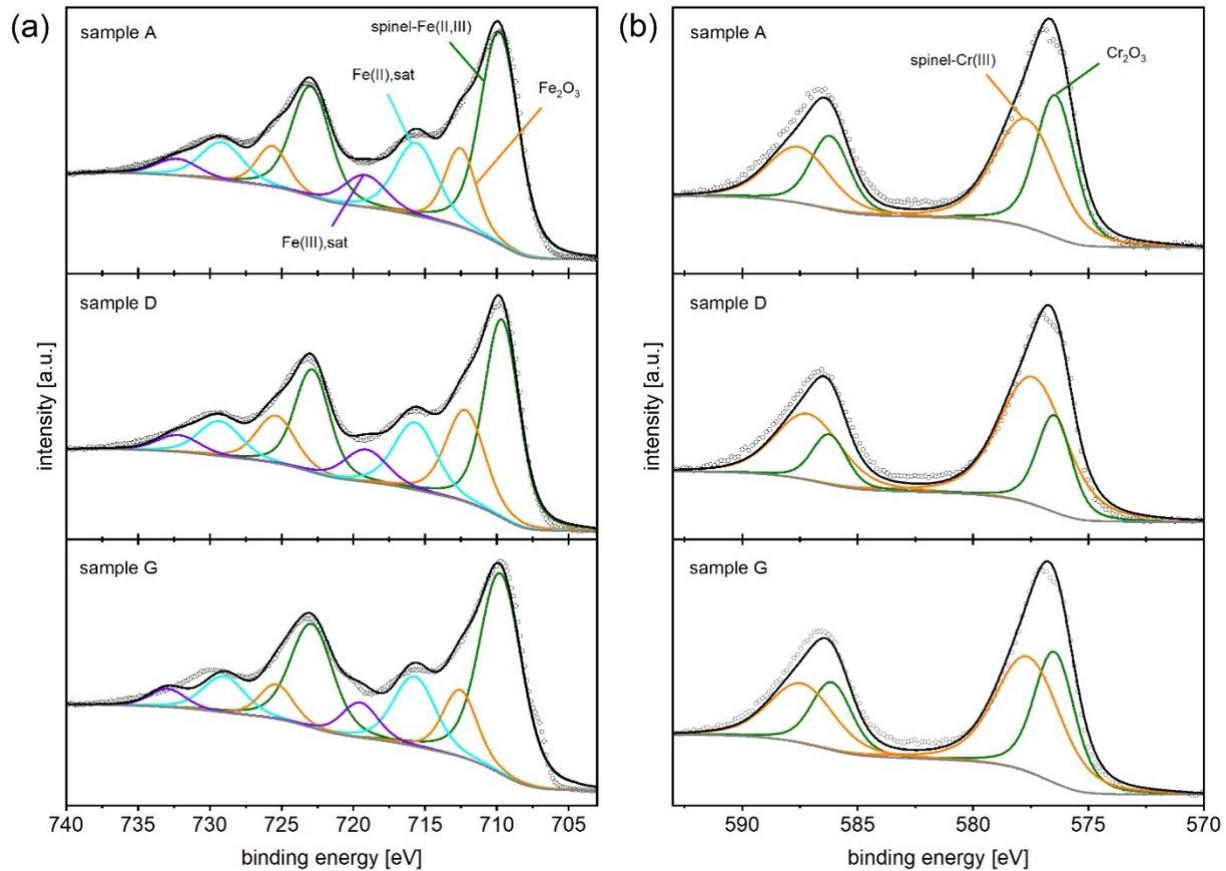

*Figure 3: XPS spectra of (a) Fe 2p and (b) Cr 2p. The symbols represent the actual measurement, whereas the colored lines are the corresponding fits. The black and grey lines are the fitting envelope and the background, respectively.*

In the DLW process, the beam diameter at the focus plane was 60 µm, resulting in a diameter of the oxide formation area increasing from 55 to 85 µm, as the scanning speed was decreased from 120 to 10 mm/s. Since the oxide layer grows upon heat accumulation on the substrate, it was observed that for all scanning speeds the energy input was high enough to melt, and therefore to erase, the underlying DLIP features in the process. To combine the structural coloration with the microtexture fabricated by DLIP, the DLW hatch distance was set larger than the spot diameter, so that gaps between two adjacent DLW-lines (see Figure 1d) were possible. To fix a distance between two colorized lines of 15 µm, the hatch distance was adjusted from 65 to 90 µm for scanning speeds decreasing from 120 to 10 mm/s, resulting in oxide lines width from 50 to 75 µm, respectively.

On the final laser-treated plate shown in Figure 2b, it can be observed that the lowest scanning speed of 10 mm/s (samples A and A[+]) yielded a deep blue color shifting over magenta, red, orange, yellow to light blue with increased scanning speed. This is true for DLIP-structured and featureless areas with the consideration that structured areas tended to appear more saturated. The characteristic iridescent effect, or rainbow coloration, typically seen on periodic surface microfeatures [54] can be observed on the area only structured by DLIP and labeled as R[+] at a tilted angle (Figure 2c). Although at high enough light intensities, like direct sun exposure, the rainbow effect was visible in all DLIP-treated samples, at indoor lighting the holographic effect was hardly visible. The label R stands for the reference flat and untreated steel surface.

Figure 4 shows confocal images of exemplary selected pairs of samples with the same colorized fields, i.e. same DLW parameters. The image with the label R corresponds to the untreated steel surface, whereas the label R+ corresponds to the DLIP-structured surface. In the structured reference R+, the DLIP pattern with a periodicity of 1 µm can be clearly observed. The topography images of the colorized samples (pairs A/A+ and H/H+) feature periodic grooves alternating between the DLW-produced lines and the underlying areas. Furthermore, the images corresponding to each color pair taken with the low magnification look remarkably similar. However, when observing the topography images in the insets with higher magnification, the DLIP microtexture becomes evident in the gaps between adjacent DLW-lines of samples A+ and H+. In contrast, in samples A and H the area between the DLW ridges is smooth.

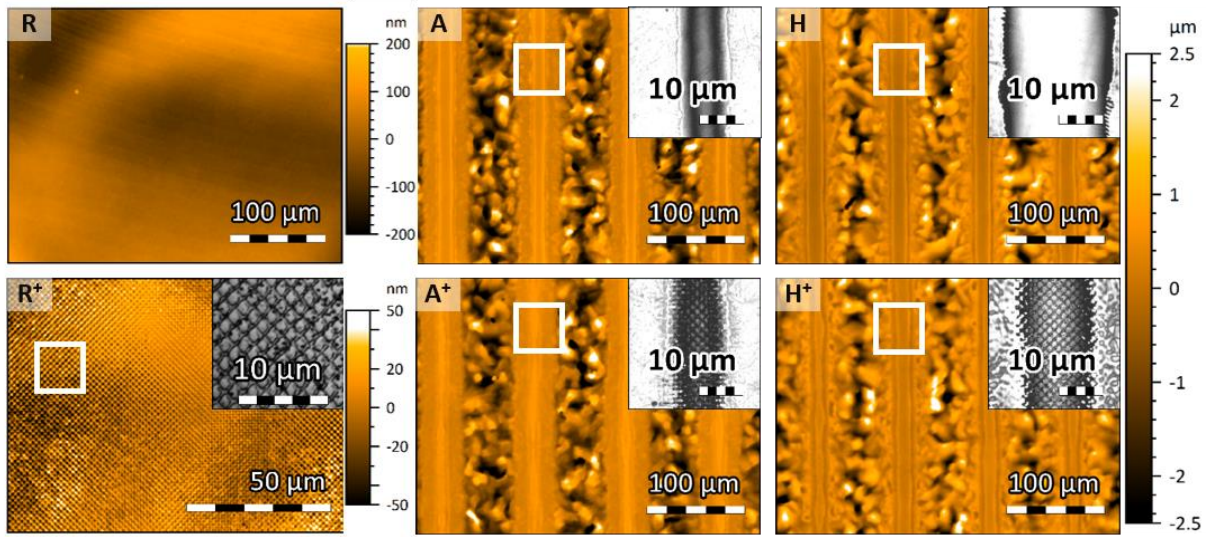

Figure 4: Confocal images of exemplarily selected samples, with non-colorized references R (untreated) and R+ (DLIP-structured). Colorized samples A and H (featureless), and A+ and H+ (DLIP-structured).

It was observed that for the same DLW parameters, the spacing between DLW-lines was overall narrower in the pre-structured samples than on featureless surfaces. A probable explanation for this finding is that the absorption of the laser radiation was slightly higher in the DLIP-patterned samples due to the higher initial roughness which in turn increased the heat accumulation and the lateral heat-affected zone. Since the hatch distances were chosen to target 15 µm gaps on the structured area, the gaps in the featureless samples were approximately 2 up to 6 µm wider.

The change in the perceived color between the structured and featureless samples irradiated with the same DLW parameters could be explained by a difference in the oxide thickness. As the laser absorption depends on the material as well as on the initial surface roughness [55], the DLIP-patterned samples probably had a higher absorption leading to a larger heat-affected zone during the DLW process. This effect was already observed in the insets of Figure 4. In the same direction, as the DLIP-patterned surfaces absorbed more light than the corresponding flat surfaces, the appearance of the structured samples looks darker.

## 3.2 Diffraction analysis of laser-treated samples

Analyzing the diffraction patterns captured with the diffractive measurement system can not only reveal the presence of periodic microstructures on a surface, but also the periodicity and geometrical arrangement. This can be easily confirmed by the images taken from the plain reference R and the DLIP-structured reference R+ shown in Figure 5. While the image corresponding to R is simply the specular reflection by a polished surface, a characteristic diffraction pattern of four-beam DLIP structures can be seen for R+ [56]. The slightly right and upward shifted spot near the center of every image is an artifact from the utilized lambda-quarter-plate. Higher-order diffraction peaks appearing near the edges of the images became distorted and stretched toward the center because of spherical aberrations, as it is clearly visible in R+. Exemplary images corresponding to the featureless colorized samples A and H are shown in the top row of Figure 5. The horizontal stretching of the diffraction patterns of samples A and H can be associated with the light scattered from the DLW-lines, which are aligned vertically (relative to the CCD images of Figure 5). For structured samples R+, A+ and H+, the images in the bottom row are split so that the left side is shown as captured and the right side is enhanced to increase the contrast. In these samples, all the first and partially the second diffraction orders were collected with the optical setup, further proving the existence of periodic surface microfeatures on the samples.

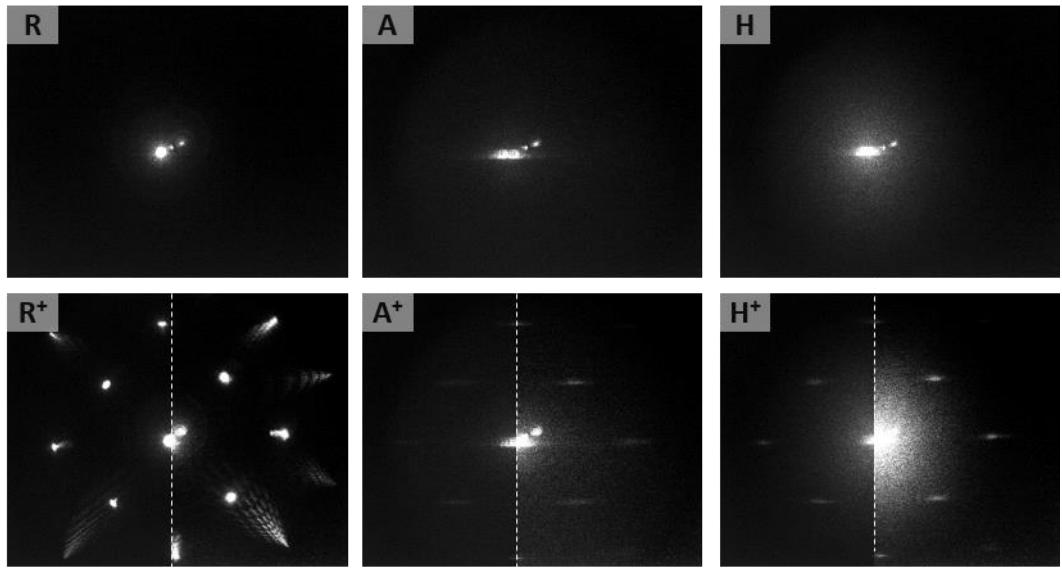

*Figure 5: Images captured with the diffraction measurement system of selected samples: Non-colorized references R (flat steel) and R+ (DLIP-structured); colorized samples by DLW treatment on flat steel A and; and colorized surfaces A+ and H+ by DLW on pre-patterned surfaces by DLIP.*

Due to the small DLIP-structured area within the 15 µm gaps of the colorized samples, the diffraction peaks are significantly less intense than the fully DLIP-treated area R+. The horizontal stretching of the zero and higher-order peaks observed for the A+ and H+ samples can be attributed to the convolution between the diffraction patterns associated with the DLIP features and those from the DLW grooves.

### 3.3   Spectral characterization and color perception

The photograph of the structured samples shown in Figure 2b provides an impression of the achievable color palette produced with the laser processes. To quantitatively characterize the colors of the oxidized samples with and without the DLIP texture, the global reflectance in the visible and near-infrared spectrum was measured for each sample. Figure 6 shows exemplarily the reflectance for featureless and DLIP-structured pairs A/A+, D/D+, H/H+ and L/L+. Recalling Table 2, these sets of experiments correspond to an increasing DLW scanning speed from 10 mm/s (samples A and A+) to 120 mm/s (L and L+). In samples A/A+, the spectral reflectance presented a valley at 620 nm and a peak at 830 nm. As the scanning speed increases, the heat accumulation decreases influencing the growth and final thickness of the oxide layer on the steel surface [57]. Therefore, the interference condition in the oxide cavity changes as a function of the scanning speed as can be observed for the reflectance measurements in Figure 6. Particularly, the valley at 620 nm and peak at 830 nm observed for samples A and A⁺ were shifted to shorter wavelengths as the scanning speed increased. From the photograph shown in Figure 2b, pair A/A⁺ had a purple-blue coloration that can be correlated to the reduced reflectance in the green-orange (550 to 660 nm) spectrum. Likewise, the pink-magenta color of pair D/D⁺ can be ascribed to the higher reflectance for wavelengths longer than 650 nm. In the case of the H/H⁺ pair, its yellow coloration can be associated with the broad peak in the spectral range 550 – 650 nm, whereas the light blue color seen in the L/L⁺ pair is attributed to the higher reflectance for wavelengths between 400 nm and 550 nm.

For the DLIP-structured samples, the measured reflectance was in general slightly lower than in the only-DLW processed samples. Furthermore, peaks and valleys of the reflectance spectra were shifted to longer wavelengths on the DLIP-structured samples by 20 to 50 nm compared to the featureless counterpart. This can be explained by a higher absorption of the laser radiation on the DLIP-treated samples, which could increase the heat-affected zone and induce a thicker oxide layer.

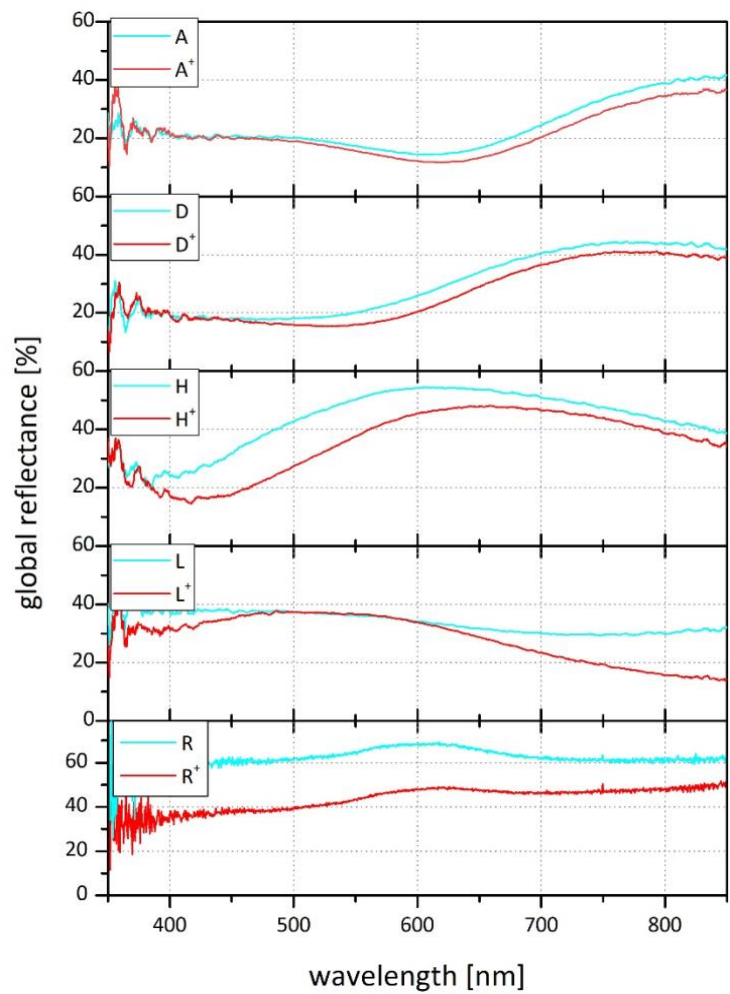

*Figure 6: Measured global reflectance over the visible spectrum for sample pairs A/A⁺, D/D⁺, H/H⁺ and L/L⁺ and reference pair R/R⁺.*

For each spectrum, also the corresponding CIE L*a*b* color space was calculated to compare perceptual differences in more detail. Before evaluation, these color space values were recalculated to a scenario with 'normal daylight' illumination as stated in the methods section. The corresponding CIE L*a*b* values of the samples are shown individually (L*, a* and b*) over the cumulated laser fluence of the DLW treatment (see Table 2) in Figure 7a for the featureless (cyan dots) and DLIP-structured samples (red squares). In Figure 7a the samples processed with the lowest scanning speed are indicated by a blue dot (featureless) and black square (DLIP-structured). It can be seen that the three-color space variables for the DLW-only and DLIP-processed samples showed similar trends as the cumulated fluence increased. The lightness L* of colorized pairs is similar (< 5 absolute) for cumulated fluences between 113 and 255 J/cm², although outside this range the relative difference in L* can reach up to 10 (absolute). The chromaticity parameters of both sets of samples showed different trends. For instance, for fluences lower than <350 J/cm² the difference in the a* parameter between both sets of samples was very low (<4 absolute), but for fluences higher than 350 J/cm² the difference became significant, implying a distortion of the perceived color. In turn, the differences in the b* parameter became evident (>9 absolute) only in the low fluence regime (< 146 J/cm²).

In Figure 7b, the featureless and DLIP-structured samples with their CIE L*a*b* values are plotted in a 3D representation of the color space. The blue dot (featureless) and the black square (with DLIP structures) indicate the samples with the highest cumulated fluence. For both sets of samples, the color pathway tends to form a closed loop as the cumulated fluence decreases, suggesting that the interference condition on the oxide cavity is similar for the lowest and highest cumulated fluences.

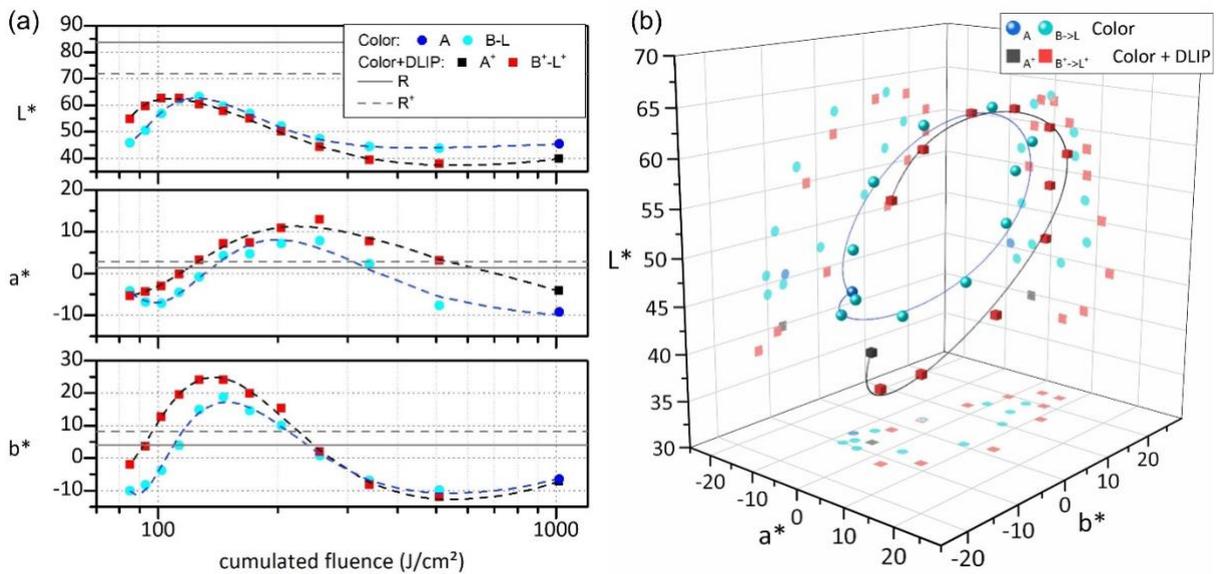

*Figure 7: CIE L*, a* and b* values as a function of cumulated fluence after the DLW process (a), with cyan circles representing featureless (A to L) and red squares standing for DLIP-structured (A⁺ to L⁺) samples; representation of distribution within the CIE L*a*b* color space (b) of samples A to L and A* to L⁺. Lines are guides to the eye and the blue dot and black square represent the samples A and A⁺, respectively.*

The color difference *dE* between two arbitrary samples *i* and *j* can be calculated according to Equation 1 [21]:

$$dE = \sqrt{(L_i^* - L_j^*)^2 + (a_i^* - a_j^*)^2 + (b_i^* - b_j^*)^2}. \qquad (1)$$

In theory, a value *dE* = 1 represents the smallest difference that the human eye can discriminate, however, values of *dE* only higher than 4 can typically be identified by the average observer and still values around 7 can be considered as acceptable in the print industry [20].

Figure 8a shows the color difference *dE* for all produced pairs as a function of cumulated fluence of the DLW process. Besides, the L*a*b* values were recalculated to sRGB and shown in Figure 8b as a digital representation and for providing an impression of the color difference between the pairs. In Figure 8a it is shown that the color difference *dE* between colorized pairs, i.e. with and without the

DLIP texture, lie below 15 for all pairs produced in this work. Although some sample pairs, e.g. A/A+, B/B+ or K/K+, have a *dE* > 10 and can be easily discriminated (see Figure 8b), the color difference *dE* of other pairs, like E/E+, F/F+ or G/G+, was below 7 (dotted horizontal line) and harder to identify. Neither the reference (R) nor the DLIP-structured surface (R+) was included in Figure 8, because they are not intended to be used as structural colors.

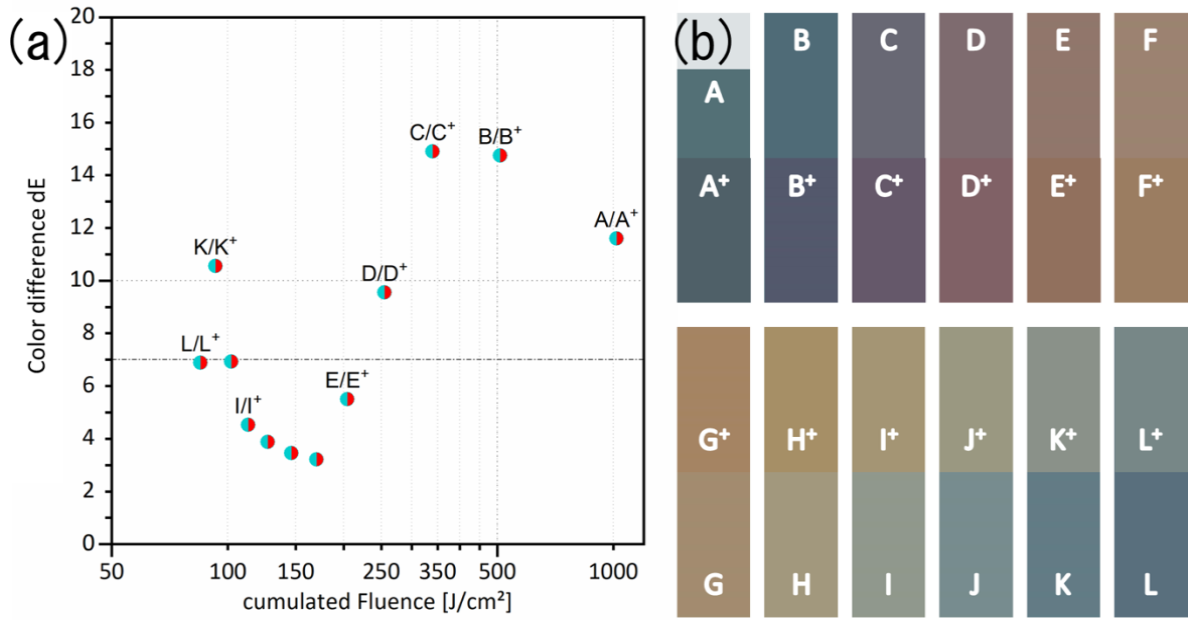

Figure 8: Calculated color difference dE over cumulated fluence for sample pairs A/A+ to L/L+ (a); sRGB value representation of the samples calculated from CIE L*a*b values (b).

A demonstrator showing 6 of the structural colors with and without underlying DLIP structures was produced on an 80 mm × 60 mm stainless steel plate. Figure 9a shows a photograph of the laser-colorized plate displaying the so-called "Rubik" cube. To obtain the different colors, the DLW process with the parameters shown in Table 2 was used. Some of these color fields were previously treated with two-beam DLIP, similarly as in the experiments described above. To this end, two of the four sub-beams were blocked in the DLIP head, while the remaining two beams were overlapped on the sample surface, forming the interference pattern. For this demonstrator, the spatial period was fixed to 3 µm, an overlap of 60 % was used with a spot diameter of 50 µm, a pulse energy of 25 µJ and a fluence of 1.3 J/cm$^2$ at a repetition rate of 10 kHz was set.

Exemplarily, the "○" symbol in Figure 9a indicates two featureless fields, whereas the "+" symbol marks two DLIP-structured areas. In this case, the DLW structuring direction was rotated, as shown in the microscopy image of Figure 9b, which yielded inclined and elongated diffraction lines in the CCD images collected with the monitoring system (see Figure 9c). Also the image in Figure 9b shows the orientation of the DLIP-engraved grooves (red lines). Therefore, in the colorized field including DLIP-structures ("+"), the first-order diffraction peaks convoluted with the DLW inclined diffraction lines are arranged horizontally. Finally, the photographs of Figure 9d display the diffraction patterns projected on a screen of a featureless (left) and DLIP-structured (right) upon illuminating the sample with a laser pointer. This demonstrator shows that combining DLW and DLIP processing is possible to hide information, in the form of diffraction patterns, within colorized surfaces. The hidden information can serve as a template for hard-to-imitate anti-counterfeit measures for protecting original goods.

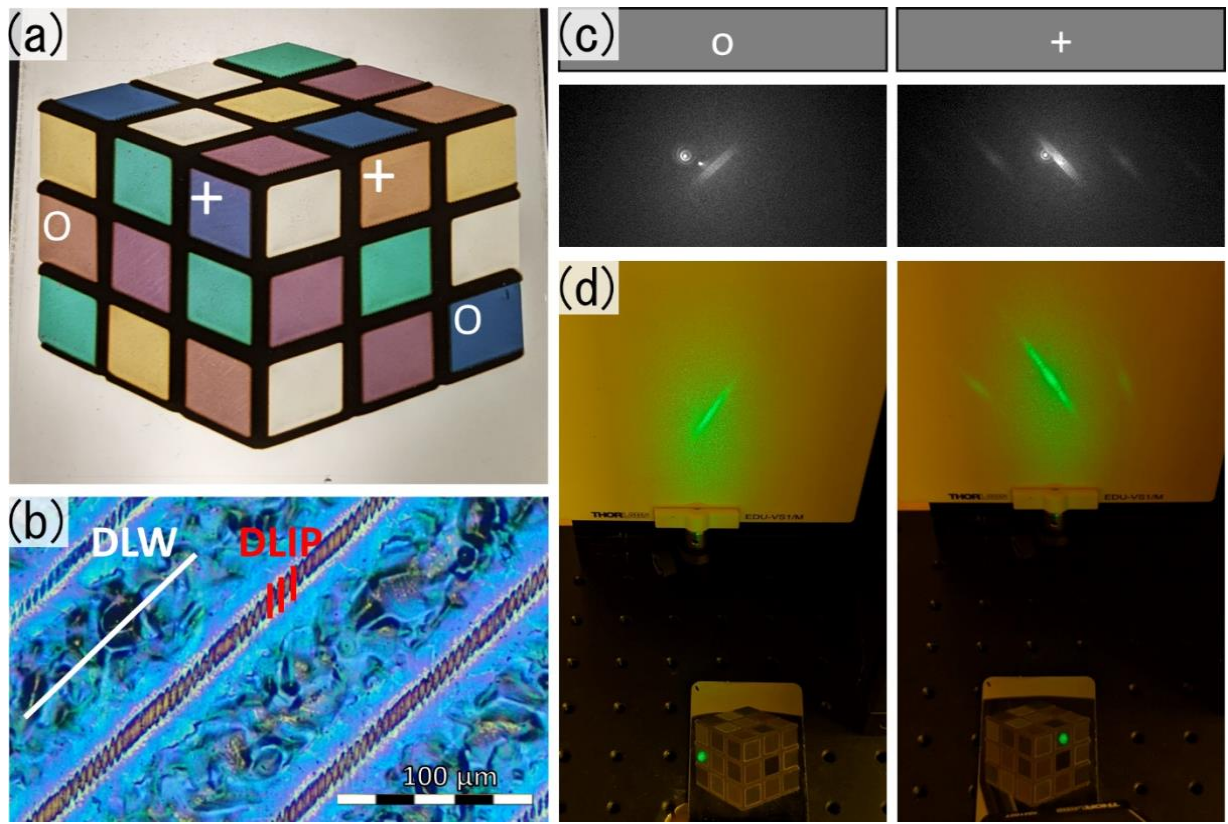

*Figure 9: Photograph of colorized motive with hidden security features (a); optical microscopy image of the DLW-treated lines and DLIP-structure orientation (white and red lines, respectively) (b); monitoring system images taken from a field with DLIP-structures "+" and from a featureless field with the same color "○" (c). Photographs of diffraction patterns projected on a screen from the corresponding fields labeled as "○" (left) and "+" (right) and upon being illuminated with a laser pointer (d).*

## 4 Conclusions

In this work, the techniques of laser-based colorization due to the growth of a oxide layers and direct laser interference patterning to introduce periodic microstructures on a metal surface were combined. It was shown that the DLW-treatment on flat and DLIP-structured areas yielded a colorful palette. Furthermore, the DLIP structures could be preserved within the gaps of the DLW lines, which was verified by topographical confocal measurements as well as diffraction analysis with an optical system. The comparison of spectral reflectance data, and therefore the CIE L*a*b* color space values, between the featureless and DLIP-structured samples yielded similar trends upon varying the DLW scanning speed. Furthermore, the CIE L*a*b* color difference for each pair of featureless and DLIP-structured samples did not exceed 15 (absolute) and the best sample pairs had a color difference of just ~3.5 (absolute). These differences in color perception can be further reduced by optimizing the DLW parameters for each targeted color to make featureless and structured areas even harder to distinguish from each other.

A demonstrator was manufactured to confirm the functionality of the presented approach. Here, some colorized fields have hidden diffractive microstructures which can be easily identified by illuminating them with a laser pointer. In this way, security features, for example as an anti-counterfeit measure, can be discreetly embedded in the surface of a product.

Other applications for this approach could be decorative colorized surfaces featuring an iridescent diffraction effect. To this end, the DLIP-structured area must be increased and the microtexture optimized. Additionally, special crafted DLIP-structures with a combination of pattern shapes and periods can be devised to generate a unique diffraction pattern from a colorized surface without the rainbow effect.

# 5 Acknowledgments

This work was carried out in the framework of the Reinhart-Koselleck project (LA2513 7-1), which has received funding from the German Research Foundation (German: *Deutsche Forschungsgemeinschaft* DFG). The work of A.L. is also supported by the German Research Foundation (DFG) under the Excellence Initiative program by the German federal and state government to promote top-level research at German universities. M.S. acknowledges the support of the Alexander von Humboldt Foundation. Y. V. thanks the European Research Council (ERC) for support under the European Union's Horizon 2020 research and innovation program (ERC Grant Agreement No. 714067, ENERGYMAPS).